# A PAIR OF QUANTA CANNOT BE WED

Nick Herbert
quanta@cruzio.com

ABSTRACT: Wooters, Zurek and others have shown that "A Single Quantum Cannot Be Cloned". The reason is two-fold: 1. A quantum cloner would permit FTL signaling; 2. A quantum cloner would violate the linearity requirement for quantum superposition. I present here a similar proof that two arbitrary quantum states cannot be universally welded together to produce a double quantum state. In particular, opposite polarization states cannot be perfectly merged. This paper closes another FTL loophole and discloses a new law of nature: Perfect quantum weddings are not possible.

INTRODUCTION
In the canonical two-photon EPR setup, each photon (A and B) of a pair of correlated photons in the quantum state:

$$\Psi(A,B) = |H(A)\rangle|H(B)\rangle - |V(A)\rangle|V(B)\rangle$$

which is the same as

$$\Psi(A,B) = |R(A)\rangle|R(B)\rangle - |L(A)\rangle|L(B)\rangle$$

is sent from Source O to two distant observers Alice and Bob.

If Alice chooses to measure Plane Polarization (PP = Horizontal (H) and Vertical (V) eigenfunctions) then Alice will observe a random sequence of H and V photons which I call "PUP light" for "Plane Unpolarized light". Because of the perfect correlation of the EPR state, if Bob also choses to measure PP light, Bob will record an identical sequence of PUP photons.

On the other hand if Alice chooses to measure Circular Polarization (CP = Right (R) and Left (L) eigenfunctions) she will observe a random sequence of R and L photons which I call "CUP light" for "Circularly Unpolarized light". If Bob chooses to measure CP light he will record a random sequence of CP photons that is identical to Alice's sequence.

To use this EPR setup for FTL signaling, Bob has only to discover a way of distinguishing a random sequence of PP photons (PUP light) from a random sequence of CP photons (CUP light). A clue that this distinction may be difficult to make is the fact that quantum theory does not recognize the difference between CUP and PUP light--describing them both by the same diagonal density matrix.

On the other hand an important physical difference seems to exist between these two sequences. For example, Alice can predict with 100% certainty the outcome of certain of Bob's single-photon measurements which suggests that there is more information contained in Bob's photon sequence than can be captured in the conventional density matrix description. If Bob can access this "hidden information" then he can learn which polarization basis Alice is using and read off her instant distant actions "faster than light".

What prevents Bob from discovering the state of his single photon is the general notion that the state of a single quantum cannot be observed. Although this belief is widely shared among the physics community, I am not aware of an impossibility proof of this conjecture. For example, the simultaneous knowledge of position and momentum is impossible because quantum mechanics does not permit states to exist that are simultaneously eigenfunctions of these two observables.

One cannot observe the position and momentum of a single quantum state because such observables simply don't exist  However such a proof for the unobservability by Bob of Alice-produced CUP light would not seem viable, for Alice knows with certainty the hidden properties of Bob's photons. Unlike simultaneous position and momentum, Bob's photons'  polarizations objectively exist. Why can't Bob measure them?

SINGLE-PHOTON MEASUREMENT
In the absence of a proof that "a single quantum cannot be known", some physicists have been motivated to devise schemes by which the secret knowledge encoded in Bob's photons might be made manifest. One such scheme was my FLASH (2) proposal which envisioned a device that could "clone" Bob's photons which would allow several measurements to be carried out on the resultant identical states and hence reveal the hidden objective polarizations of Bob's photons.

FLASH was refuted by Wooters, Zurek (1) and others who showed that a perfect cloner could indeed be built that would clone two orthogonal states, but a universal cloner (that could clone any unknown state) would violate the quantum principle of linear superposition.

Oddly enough, on the same page of Nature in which Wooters and Zurek's refutation appeared, a note by Leonard Mandel (3) described a simple two-atom universal cloner which, in retrospect, turns out to be the optimum cloner allowed by quantum theory. Mandel's cloner takes any polarization state |Z> and outputs the state |ZZ>. However, a certain fraction of times Mandel's device will produce the state |ZZ#> where Z# is the polarization orthogonal to Z. In Mandel's device the fraction of times that the cloner produces the "wrong polarization" is 1/3.

Let's see how Mandel's optimum cloner will work on Bob's photons to aid Bob's attempt to distinguish PUP from CUP light.

Inputting PUP light--a random sequence of |H> and |V> photons:
An |H> photon turns into a |HH> state twice and once into an |HV> state.
A |V> photon turns into a |VV> state twice and once into an |HV> state.

$|H\rangle$ & $|V\rangle$ ---> 2 $|HH\rangle$ & 2 $|VV\rangle$ & 2 $|HV\rangle$   Operation of Mandel cloner

The end result is an incoherent sum of $|HH\rangle$ & $|VV\rangle$ & $|VH\rangle$ biphotons in equal proportions. But this state can be shown to be mathematically equivalent to: $|RR\rangle$ & $|LL\rangle$ & $|RL\rangle$ which is the output of the Mandel cloner for CUP light.

The Mandel cloner applied to CUP or to PUP light produces an identical output. A Mandel cloner takes unpolarized "one-light" and turns it into unpolarized "two-light".

Note that unpolarized two-light consists of an equal intensity of each of the three possible polarization combinations. Any other proportion of intensities is not unpolarized two-light and can in principle be experimentally distinguished from it.

QUANTUM WEDDING
Unpolarized one-light of the PUP type consists of a random sequence of $|H\rangle$ and $|V\rangle$ photons (whose identities are known to Alice but not to Bob). Suppose however that Bob possesses a device that merges each photon with the one that follows it in a "quantum wedding". Bob might do this by gaining knowledge of the emission time of each photon and simply delaying a particular photon in some exotic medium till its laggard follower catches up and superposes itself on the earlier photon. Thus if the early photon is of the $|H\rangle$ type and the subsequent photon is of the $|V\rangle$ type, the "quantum wedding" will produce a biphoton of the type $|HV\rangle$.

It is easy to see that consistent application of this merge operation to PUP light will produce an incoherent sequence of biphotons with the following structure:

QUANTUM WEDDING (PUP) = $|HH\rangle$ & $|VV\rangle$ & $|HV\rangle$ & $|VH\rangle$ = "Fat PUP"

On the other hand application of this hypothetical merge operation to CUP light will produce an incoherent sequence of biphotons with the following structure:

QUANTUM WEDDING (CUP) = $|RR\rangle$ & $|LL\rangle$ & $|RL\rangle$ & $|LR\rangle$ = "Fat CUP"

where "&" indicates incoherent addition.

Unlike the Mandel cloner which produces the same kind of unpolarized two-light for both PUP and CUP inputs, the quantum wedding produces two experimentally different outputs for PUP and CUP inputs. In other words:

> Fat PUP is not equal to Fat CUP

For example, these two kinds of light applied to an H/V beamsplitter produce decidedly different outcomes.(see below). This difference could be used by Bob to decode Alice's hidden message and hence achieve EPR-mediated communication faster than light. In order to prevent this eventuality from occurring in nature, I propose a new quantum commandment:

"A pair of quanta cannot be wed."

NO QUANTUM MARRIAGE MILLS

I have shown above that given the ability to "marry" H, V, R and L photons, one can use the EPR setup to send signals FTL. If one is able to wed consecutive photons in a CUP or a PUP beam, one produces not CUP or PUP two-light but "Fat" CUP or "Fat" PUP two-light with an excess of |RL> and |HV> photons respectively. If we can produce "Fat" unpolarized light of the CUP and PUP variety, this light can be used to experimentally distinguish the polarization of Bob's photons which were distantly produced by Alice.

To prevent FTL signaling, a universal quantum marriage mill must be impossible. Such a marriage mill would possess the following properties.

$|H>_1 |H>_2 A_o \longrightarrow A_H |2H>_3$ $\qquad$ $|R>_1 |R>_2 A_o \longrightarrow A_R |2R>_3$
$|V>_1 |V>_2 A_o \longrightarrow A_V |2V>_3$ $\qquad$ $|L>_1 |L>_2 A_o \longrightarrow A_L |2L>_3$
$|H>_1 |V>_2 A_o \longrightarrow A_M |HV>_3$ $\qquad$ $|R>_1 |L>_2 A_o \longrightarrow A_E |RL>_3$
$|V>_1 |H>_2 A_o \longrightarrow A_N |HV>_3$ $\qquad$ $|L>_1 |R>_2 A_o \longrightarrow A_F |RL>_3$

The marriage mill begins in its initial state Ao with two photons A, B in the product state $|A>_1 |B>_2$. The mill unites these photons in the wedded state $|AB>_3$ while the mill itself changes into the state $A_X$.

The wedding of two like polarized photons can be easily accomplished by inserting them into two ports of a 50/50 beamsplitter. The Hong Ou Mandel Effect (4) guarantees that the photons will exit united in the same output port of the beamsplitter. How to wed states of unlike polarizations is more problematic but if it is at all possible, some clever experimentalist will invent a way to accomplish it.

As in the proof of the no-cloning theorem I use the definitions that connect H,V and R,L polarization states and require that the equations satisfy the linear superposition property of quantum mechanics. Linearity constrains the coefficients of the marriage mill in the following manner:

$A_H = A_V = A_R = A_L = A_S$ (same-sex wedding state)

$A_M = A_N = A_E = A_F = A_O$ (opposite-sex wedding state)

Good news. At this stage the quantum marriage mill seems entirely possible. No obvious contradictions arise. But linearity also compels one additional relationship:

$$(A_O)^2 = 1/2(A_S)$$

This relationship requires that a same-sex marriage must be twice as likely as an opposite sex marriage. If the marriage mill glues together like photons with 100% efficiency then unlike photons will be glued together with only 50% efficiency.
In other words unlike photons cannot be reliably merged. At least half the time they must enter some other unglued state. To conserve probability (the unwed photons must go somewhere) the final state of a real marriage mill must allow for at least one additional possibility for attempted heterosexual union. For instance:

$$|H>_1 |V>_2 A_0 \longrightarrow A_O |HV>_3 + A_Z |H,V>_4$$

where $|H,V>_4$ is some state where the H, V photons are not united.

The possibility of FTL signaling via the quantum wedding of consecutive EPR photons depends crucially on producing "Fat unpolarized two-light"--that is, light with an excessive number of, say, $|HV>$ biphotons. It is not surprising to discover that quantum mechanical linearity outlaws perfect quantum weddings and that such weddings fail in precisely such a manner as to halve the number of $|HV>$ biphotons produced. Thus a real quantum marriage mill will indeed glue photons together but will never produce Fat unpolarized two-light but only ordinary unpolarized two-light for which the CUP and PUP forms are experimentally indistinguishable.

This result closes another loophole (the "glue" loophole) that might have been used to exploit the EPR effect to send FTL signals. It also uncovers a new law of nature: Perfect quantum weddings are not possible.

NICK HERBERT
FEBRUARY 2008

REFERENCES

(1) W. K. Wooters and W. Zurek, Nature **299**, 802 (1982)

(2) N. Herbert, Foundations of Physics **12**, 1171 (1982)

(3) L. Mandel, Nature **299**, 802 (1982)

(4) C. K. Hong, Z. Y. Ou and L. Mandel, Phys. Rev. Lett. **59**, 2044 (1987)

================================================================
TWO-LIGHT INCIDENT ON AN H/V BEAMSPLITTER
Here are illustrated various types of unpolarized two-light and their behaviors at a H/V beamsplitter. I call light with an excess of cross terms of the type $|HV>$ "Fat unpolarized light".

The Ratio R represents the relative number of photons that go into the same polarization channel [HH] or [VV] divided by the number that go into two different channels [HV]. "&". as above, indicates incoherent addition.

Plane-unpolarized (PUP) two-light = IHH> & IVV> & IHV>

Circular-unpolarized (CUP) two-light = IRR> & ILL> & IRL>

Fat Plane-unpolarized (FPUP) two-light = IHH> & IVV> & IHV> & IVH>

Fat Circular-unpolarized (FCUP) two-light = IRR> & ILL> & IRL> & ILR>

| TYPE OF LIGHT | [HH] | [VV] | [HV] | R |
|---|---|---|---|---|
| PUP two-light | 1 | 1 | 1 | 2 |
| CUP two-light | 1 | 1 | 1 | 2 |
| FPUP two-light | 1 | 1 | 2 | 1 |
| FCUP two-light | 3 | 3 | 2 | 3 |

Note that PUP and CUP light produce identical outputs but Fat PUP and Fat CUP produce outputs different both from one another and from PUP and CUP two-light.
===================================================================